\documentclass[prl,twocolumn,showpacs]{revtex4}
\usepackage{graphicx}
\usepackage{dcolumn}
\usepackage{bm}

\begin{document}

\title{Radiation induced zero-resistance states: a dressed electronic structure effect}

\author{P.~H.~Rivera} 
\affiliation{Consejo Superior de Investigaciones, Facultad de Ciencias
F\'{\i}sicas, Universidad Nacional Mayor de San Marcos, Lima, Per\'u} 
\author{P.~A.~Schulz}
\affiliation{Instituto de F\'{\i}sica `Gleb Wataghin', Universidade Estadual de Campinas,
13083--970, Campinas, S\~ao Paulo, Brazil}

\begin{abstract}
Recent results on magnetoresistance in a two dimensional electron gas under crossed
magnetic and microwave fields show a new class of oscillations, suggesting a
new kind of zero-resistance states. 
A complete understanding of the effect is still lacking.
  We consider the problem
from the point of view of the electronic structure 
 dressed by photons due to a in plane linearly
polarized ac field. The dramatic changes in the dressed electronic
structure lead to a interpretation of the new magnetoresistance oscillations as a 
persistent-current like effect, induced by the radiation field.
\end{abstract}

\maketitle

  A new class of low temperature non-equilibrium zero-resistance states
   (ZRS) have recently been identified by Mani and coworkers in
   irradiated  quantum Hall systems based on GaAs/AlGaAs
   heterostructures\cite{mani}.  The effect was confirmed in ultra
   high mobility GaAs/AlGaAs  quantum wells by Zudov et
   al.\cite{zudov}, who cited the phenomena as evidence for a new
   dissipation less effect in 2D electronic transport.  Such ZRS are
   induced in the two dimensional electron gas (2DEG) by
   electromagnetic-wave excitation  with the ac electric field
   parallel to 2DEG, in a weak, static,  perpendicular magnetic
   field. Oscillations of the resistance  induced by microwave
   excitation, in low mobility specimens, had  been reported
   previously \cite {zudov2,ye}. There is strong experimental
   evidence that the ZRS coincide with a gap in the electronic
   spectrum,  although the positions of the extrema remains
   controversial.  Mani and coworkers \cite{mani,mani2} find
   resistance minima(maxima) at $\omega/\omega_c=\epsilon=j+1/4
   (j+3/4)$,where $\omega$  is the ac field frequency, $\omega_c$ the
   cyclotron frequency and $j=1,2,3...$ is the difference between the
   indexes of  the participating Landau levels (LLs).  Zudov {\it et
   al.}\cite{zudov} report  different periodicities for the  maxima
   and minima, with maxima at $\epsilon=j$ and minima on the high
   field  side of $\epsilon=j+1/2$. Interest in this novel ZRS has
   already produced an extensive list of preliminary results,  which
   aim to establish a theory for understanding this  remarkable
   effect\cite{phillips,girvin,andreev,anderson,shri,mikhailov,shixie,koukalov,volkov,chere,doro}.
   We show that a full understanding, however, has to consider the
   effect from the point of view of the LLs dressed by photons due  to
   the coupling to ac fields.

The experimental parameters reveal a rich physical scenario. In summary, the quantum
Hall effects are observed at high magnetic fields ($B>0.4$ T), while for low magnetic fields
 ($B<0.4$ T) new oscillations in the
magnetoresistance are observed under radiation.  Indeed, the resistance vanishes, in a
given data collection at $B \approx 0.2$ T, for a radiation frequency of $\nu \approx
100$ GHz.  At  $B \approx 0.2$ T the LL separation is $\hbar \omega_c \approx 0.35$
meV, while the microwave photons have energy of the same order, namely $h\nu \approx 0.4
$ meV.  At the measurement temperature $T=1.5$ K, LLs are still resolved, since
Shubnikov-de Haas oscillations are still seen below $B=0.2$ T in absence of radiation. On the other hand, 
these new oscillations are observable down to $B \approx 0.02$ T \cite{mani2}, corresponding to a magnetic 
length $l_c \approx 0.18$  $\mu$m and a classical cyclotron radius up to $R_c \approx 4.5$ $\mu$m 
 \cite{mani}, still small 
compared to a mean free path of $l_0 \approx 140$ $\mu$m.
  Besides that, the used  microwaves have frequencies down to $\nu = 30$ GHz \cite{zudov},
 corresponding to a wavelength 
up to $\lambda = 10$ mm, approximately a factor of 2 or 3 larger than the linear dimensions of the sample, $w$. 
Indeed, oscillations have been reported at frequencies down to 3 GHz ($\lambda = 10$ cm) 
\cite{willet}. 
A ratio $\lambda/w \approx 10$ validates a dipole approximation for the radiation-sample 
coupling.
More important is that
 the estimated power level is of $\le 1$ mW, over a
cross sectional area of $\le 135$ mm$^2$ in the vicinity of the sample \cite{mani}. 
This represents a
 field intensity of $I=7.4$ $W/m^2$, which can be related to the associated
electric field by $I = E^2_{rms}/(c\mu_0)$ \cite{halliday}:  $E_{rms} \approx 50$ V/m.
Considering the classical cyclotron radius as the 
 relevant length scale \cite{mani3}
and a frequency of 100 GHz, this leads to a
ratio $eR_cE_{rms}/(h\nu) \approx 0.35$ at $B=0.2$ T.  Such ratio, between the energy gained 
from the ac field over 
a distance corresponding to the cyclotron radius and the photon energy, 
represents already a field intensity that can not
be considered perturbatively \cite{dunlap}. 
 We address this problem  within a tight-binding approach, considering
non-perturbatively the two main ingredients of the problem:  Landau quantization 
and dressing of the electronic states by means of a coupling with the ac fields.
The dressed electronic structure shows non-trivial features which are clear signatures of the newly observed 
resistance oscillations.  

The present  model is a finite tight-binding lattice coupled non-perturbatively to an ac field by means of the
 Floquet method \cite{shirley}.  The
time-independent infinite matrix Hamiltonian obtained from transforming
 the time-dependent Schr\"odinger equation, describes
entirely these processes without any further {\it ad-hoc} hypothesis.
Therefore, the effects of an intense ac field on the electronic spectra have to be 
described by very large truncated matrix Hamiltonian. 
 Such numerical endeavour is possible by means of a renormalization
 procedure,  providing
  the spectral modulation as function of field intenstiy, as well as the
 intensity hierarchy of the quasi-energy spectra related to different photon replicas \cite{pablo}. 
The bare energy spectrum is the one of a
 tight-binding square lattice of \textit{s}-like orbitals, considering only nearest
 neighbors interaction.  The magnetic field is introduced by means of a Peierls
 substitution in the Landau gauge ${\bf A}=(0,l_1aB,0)$ \cite{sivan}.  An ac field will be
 considered parallel to one of the square sides.  Hence, the model for the bare
 electronic system coupled to an ac filed is described by the
 Hamiltonian $ H= H_o+H_{int}$, where

\begin{displaymath}
H_o=\sum_{l_1,l_2}\epsilon_{l_1,l_2}\sigma_{l_1,l_2}\sigma_{l_1,l_2}^{\dag}+{V\over2}\sum_{l_1,l_2}\Bigl[\sigma_{l_1,l_2}\sigma_{l_1+1,l_2}^{\dag}+
\end{displaymath} \begin{equation}
\sigma_{l_1+1,l_2}\sigma_{l_1,l_2}^{\dag}+e^{i2\pi\alpha
l_1}\Bigl(\sigma_{l_1,l_2}\sigma_{l_1,l_2+1}^{\dag}+\sigma_{l_1,l_2+1}\sigma_{l_1,l_2}^{\dag}\Bigr)\Bigr]
\end{equation}

and

\begin{equation} H_{int}=eaF\cos \omega t\sum
_{l_1,l_2}\sigma_{l_1,l_2}l_1\sigma_{l_1,l_2}^{\dag}. \end{equation}

\noindent Here $\sigma_{l_1,l_2}=|l_1,l_2>$, $\sigma_{l_1,l_2}^{\dag}=<l_1,l_2|$, where
$(l_1,l_2)$ are the $(x,y)$ coordinates of the sites. The 
phase factor $\alpha$ is defined as $\alpha=\Phi/\Phi_e$, where 
$\Phi_e=h/e$ is the magnetic flux quantum, and $\Phi=a^2B$ is the magnetic flux per 
unit cell of the square lattice. 
 The \textit{atomic energy} will
be taken constant, $\epsilon_{l_1,l_2}=4|V|$, for all sites. 
The  hopping parameter can  emulate the
electronic effective-mass for the GaAs bottom of the conduction band, $m^*=0.067m_0$.
Since $V=-\hbar^2/(2m^*a^2)$, $V=-0.142$ eV for a lattice parameter of $a=20$ \AA.
 The ac field is defined by its frequency and amplitude, 
$\omega$ and $F$, respectively.
The treatment of the time-dependent problem is based on Floquet states $|l_1,l_2,m>$
where $m$ is the photon index.  We follow the procedure developed by
Shirley\cite{shirley}, which consists in a 
transformation of the time-dependent Hamiltonian into a time-independent infinite
matrix.  The elements of this infinite matrix are

\begin{displaymath} \Bigl[({\mathcal E}-m\hbar\omega
-\epsilon_{l_1,l_2})\delta_{l_1^{\prime}l_1}\delta_{l_2^{\prime}l_2}-{V\over2}\Bigl\{(\delta_{l_1^{\prime},l_1-1}+\delta_{l_1^{\prime},l_1+1})\delta_{l_2^{\prime}l_2}\Bigr.\Bigr.
\end{displaymath} \begin{displaymath}
\Bigl.\Bigl.+(\delta_{l_2^{\prime},l_2-1}+\delta_{l_2^{\prime},l_2+1})\delta_{l_1^{\prime}l_1}\Bigr\}\Bigr]\delta
_{m'm} \end{displaymath} \begin{equation}
=F_1l_1\delta_{l_1^{\prime}l_1}\delta_{l_2^{\prime}l_2}(\delta _{m^{\prime},m-1}+\delta
_{m^{\prime},m+1}), \end{equation}

\noindent where $F_1={1\over2}eaF$. The energy eigenvalues, ${\mathcal E}-m\hbar\omega$, are 
quasi-energies of a system dressed by photons, shifted by multiples of the photon energy,
usually called as the $m-th$ ``photon replica" of the system, which are coupled by the ac field. 
 Diagonalization of a truncated
 Floquet matrix involves dimensions given
by $L^2(2{\it M}+1)$. $L$ is the lateral size of the square lattice 
in number of atomic sites,
 while ${\it M}$ is the maximum photon index.  Since the
ac field couples a Floquet state defined by $m$ photons to states with $m-1$ or $m+1$
photons, multiple photon processes become relevant with increasing field intensity. 
As a consequence, $M$, which determines how many ``photon replicas" are
taken into account, increases with field intensity.
 A truncated Floquet matrix is a tridiagonal block matrix which contain  
$L\times L$ diagonal blocks given by $\mathsf{E}^M=({\mathcal 
E}-m\hbar\omega)\mathsf{I}+\mathsf{H_0}$ representing a photon replica with 
the matrix elements
given by the left hand side of Eq.(3).  The coupling of system with the 
intense ac
electric field is represented by the off-diagonal blocks ${\mathcal F}$, 
which are
diagonal block matrices, with the elements given by
${\mathcal F}=F_1l_1\delta_{l_1^{\prime}l_1}\delta_{l_2^{\prime}l_2}$.

The dimension of the problem can be reduced to $L^2$ 
by means of a
renormalization procedure, based on the definition of the associated Green's 
function, $\mathsf G$, where 
${\mathsf F \mathsf G = \mathsf I}$.

A detailed discussion of this method is given in a previous work \cite{pablo}. 
The final result of this renormalization of the Floquet matrix is the dressed Green's
function for one of the photon replicas, say $M=0$, and a 
quasi-density of Floquet states, $\rho({\mathcal E}+i\eta)$ can then be
obtained:

\begin{equation} \label{eq:dos} \rho({\mathcal
  E}+i\eta)=-\frac{1}{\pi}~\mathrm{Im}\left[~\mathrm{Tr}~G_{MM}~\right].  \end{equation}

The trace of the Green's operator is taken over the atomic sites basis.

A finite square lattice in the presence of a perpendicular magnetic field 
shows a rich ``quantum dot-like" spectrum, Fig.1(a), with a low magnetic flux region dominated by 
finite sample size quantization (cyclotron 
radius large compared with the linear dimensions of the sample), as well as 
bulk LLs and edge states, well defined for higher magnetic fluxes \cite{sivan}. 
We consider a $L =10a$ square lattice, a size limited 
by computacional costs,
focusing on the states collapsing into LLs for $\Phi/\Phi_e \ge 0.1$. In Fig.1(a) the lowest two 
LLs are well defined, with a ladder of edge states between them. 
The advantage 
of finite lattices is that the magnetic flux can be continuously varied,
 as in the experimental measurements, since comensurability effects are absent 
 \cite{watson}. On the other hand, the unavoidable presence of edge 
 states could hinder the interpretation of the ac fields on the bulk LLs.
  Nevertheles, ac field effects on bulk LLs and
 edge states can be distinguished, as shown below. 
 
The spectrum of the system depicted in Fig.1(a), modified by an ac field,
  is shown in Fig.1(b), 
 for a 
 photon energy $\hbar \omega = 10$ meV \cite{comment2}, which is lower than 
  the quantum-dot-like states separation at very low magnetic fluxes and much 
 lower than the LL separation at values of $\Phi/\Phi_e $ where 
 the LLs start to be well defined. In Fig.1(b) the field 
 intensity is $eaF=5$ meV, where $a \approx l_c$ at $\Phi/\Phi_e \approx 0.1$ \cite{comment}. 
 This represents already a non-perturbative field intensity
  $eaF/\hbar\omega=0.5$. A dramatic change in the quasi density 
  of states can be observed, with a coupling between different photon replicas leading 
  to a flatening of the states,
   opening of gaps in the lower part 
  of the spectrum and a coupling between edge and bulk 
  states induced by the ac field. In the energy scale of the figure $E/h\nu=1$ is the 
  separation between successive photon replicas, which are successively less intense with increasing 
  the photon index $m$.
  At higher magnetic fluxes photon replicas of the lowest bulk LL 
  can be clearly followed. 
  Increasing the field intensity  
  leads to the formation of higher order photon replicas 
  of the lowest LLs, as well as new periodic   
  structure (as a function of magnetic flux) in the quasi-energy spectrum (not shown here). 

\begin{figure}[h]
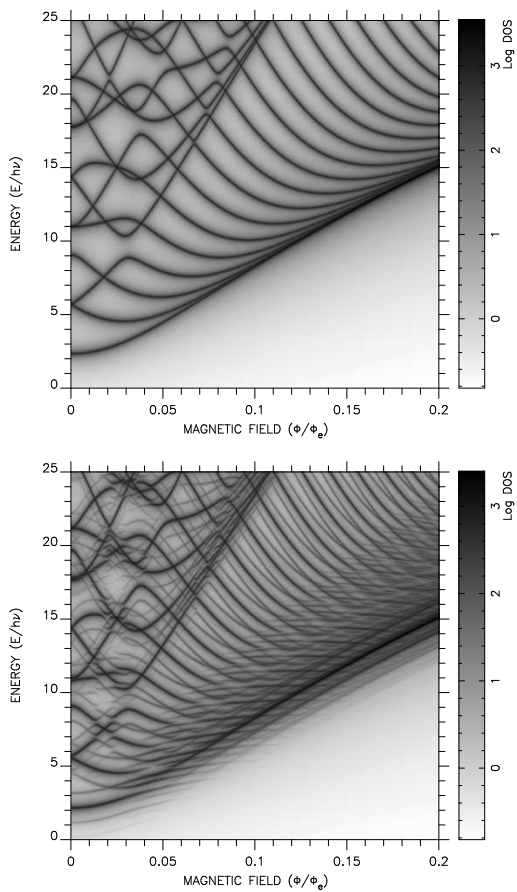

\begin{center}
\includegraphics[angle=-90,scale=0.3]{fig1a.ps}
\includegraphics[angle=-90,scale=0.3]{fig1b.ps}
\caption{Spectra of the density of states as a function of magnetic flux. Black(white) stands for 
highest(lowest) density of states. Top: spectrum for a square lattice with $L = 10a$ (see text) in 
absence of an ac field. Bottom: spectrum for the same system with an
ac field with $h\nu = 10$ meV and  
$eaF = 5$ meV.}
\label{1}
\end{center}
\end{figure}

  These results are intrinsically interesting, but we should focus on the 
  experimental conditions \cite{mani,zudov}, i.e., frequencies of the order of 
  the LL separation. In Fig.2(a) we choose $\hbar\omega = 150$ meV, namely 
  the LL separation for $\Phi/\Phi_e \approx 0.1$, a flux for which  
  the lowest LLs are already well defined. The main feature for 
  the present discussion is the avoided crossing between the second LL and 
  the first photon replica of the lowest LL (plus one photon). This 
  avoided crossing at $\Phi/\Phi_e \approx 
  0.1$ can be distinghished from other features of the spectrum in spite of the
   rather complex quasi density of states due to the ac
  filed. This anticrossing at $E/h\nu \approx 1.5$ is better observed in a zoom of the density of 
  states shown in Fig.2(b).
   An avoided crossing can also be seen  
  between the lowest LL and a first photon replica of the second LL
(minus 1 photon in this case). A crucial point is that the field intensity for the 
case illustrated in Fig.2 is $eaF=30$ meV, corresponding to $eaF/\hbar\omega=0.2$. Indeed the 
avoided crossings are already defined for $eaF/\hbar\omega=0.05$ (not shown here), a value 
one order of magnitude lower than the estimative for the actual experimental conditions discussed 
in the introduction.

\begin{figure}[h]
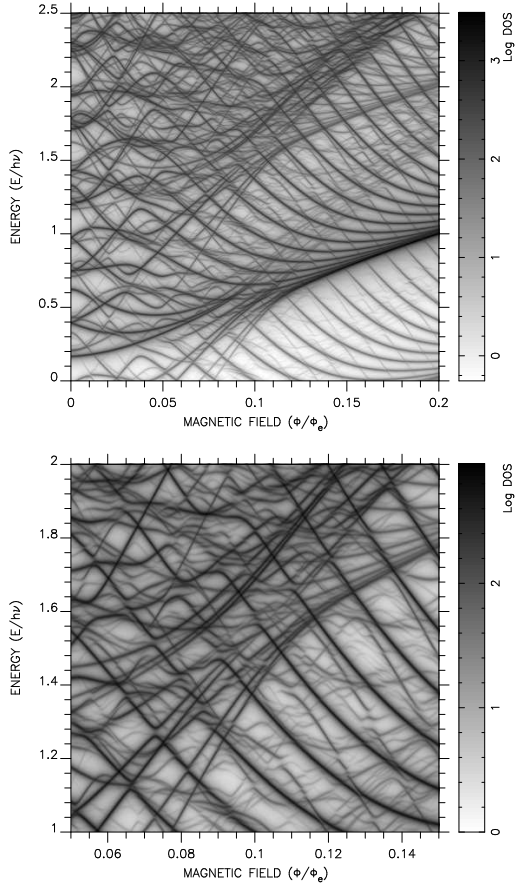

\begin{center}
\includegraphics[angle=-90,scale=0.3]{fig2a.ps}
\includegraphics[angle=-90,scale=0.3]{fig2b.ps}
\caption{Top: spectrum for a square lattice with $L = 10a$ (see text) 
with an ac field with $h\nu = 150$ meV and 
$eaF = 30$ meV. 
Bottom: zoom of the same density of states spectrum shown in the top pannel in the range of one 
of the avoided crossings at $E/h\nu \approx 1.5$ and $\Phi/\Phi_e \approx 0.1$.}
\label{2}
\end{center}
\end{figure}

Having in mind such anticrossings in the density of states, 
induced by the ac field, one can figure out a complete picture of the 
phenomenon in the sketch depicted in Fig.3. Here we represent photon replicas 
  of LLs in absence of ac field coupling. The energies are given by $E_{n,m}=e_n\pm mh\nu$, where 
  $e_n=(n+1/2)\hbar\omega_c$ are the LL energies, while $\pm m$ indicates the replicas 
  obtained by adding/subtracting $m$ photons. Only 5 LLs are show for sake of clarity. On the 
  other hand, only $m=0,1$ are considered, since only the couplig between ``nearest-neighbor 
  replicas" should be important at the field intensities considered, although higher order 
  effects are expected for even higher field intensities. Notice that the crossings between 
  $\Delta m=\pm 1$ occur only at $\omega/\omega_c = j$ (indicated by dashed vertical lines)
  The crossing of these 
  LLs become anticrossings by turning on the ac field and the anticrossings would lead the 
   a modulated 
  spectrum with the periodicity given by $\omega/\omega_c = j$. 
  The numerical calculation illustrated in Fig.2 corresponds to the last crossing 
  in Fig.3 at $\omega_c=1.0$. Oscillatory behavior of $dE/d\Phi$ are therefore 
  expected only for lower magnetic fluxes. Oscillations at higher magnetic fluxes 
  would be higher order effects ($\Delta m=\pm 2$). Such oscillating 
  spectrum resembles the spectrum of a quantum ring pierced by 
  a magnetic flux \cite{buttiker}, which reveals a persistent current. The present results 
  indicate that the recently observed ac field induced oscillations in the magnetoresistance are 
  {\it ac field induced  persistent current-like effects}. A connection between 
  static magnetic 
  flux and intense ac field effects has been recently suggested for quantum ring structures 
  \cite{vidar}. 
  
\begin{figure}[h]
\begin{center}
\includegraphics[angle=-90,scale=0.4]{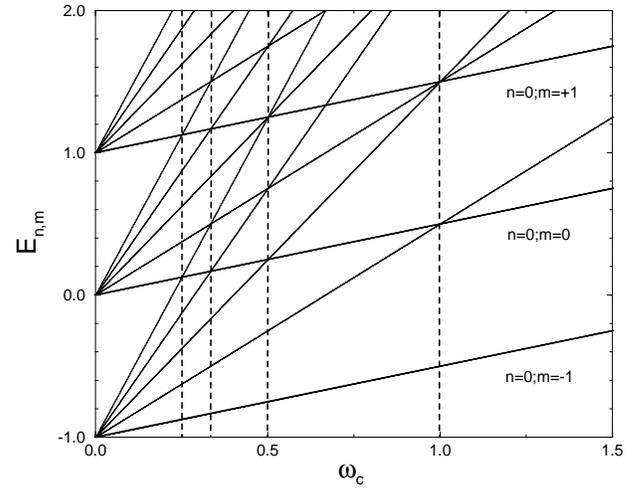}
\caption{$E_{n,m}=(n+1/2)\hbar\omega_c\pm mh\nu$ (see text), 
as a function of the cyclotron frequency, $\omega_c$. Only $m=0,\pm 1$ photon replicas are 
considered. Vertical dashed lines represents $\omega/\omega_c=j$. 
The lowest LL of each photon replica is highlighted as a guide for the eyes.}
\label{3}
\end{center}
\end{figure}

 The richness of the ac field induced features on the spectrum of a 2D system 
 threated by a perpendicular magnetic field deserves further investigations. 
 One of these features, anticrossing between LL photon replicas, 
 leads to a simple interpretation of the very interesting new measurements 
 of field induced zero-resistance states. The discrepancy in the oscillation phase, as well as 
 the hudge activation energies need further investigations within the framework proposed 
 in the present work.

The authors thank R. G. Mani for fruitfull discussions and critical reading of the manuscript.
P. A. S. acknowledges financial support from the Brazilian agency FAPESP.


\begin{thebibliography}{10}
\bibitem{comment2}
Since we are limited to small systems, the magnetic fluxes at which LLs start to be well defined 
correspond to a much higher energy scale than the Fermi energy of the actual experiments.
 Nevertheless, our arguments are independent of the energy scale, since no qualitative changes 
 are expected for larger systems.
\bibitem{comment}
The magnetic length 
 is now the appropriate length scale to decide the strength of the ac field, 
 since we are looking at the lowest LLs.
\end{thebibliography}
\end{document}